\documentclass[%
 reprint,
 amsmath,amssymb,
 aps,
]{revtex4-1}

\usepackage{graphicx}% Include figure files
\usepackage{dcolumn}% Align table columns on decimal point
\usepackage{bm}% bold math
\usepackage{psfrag}
\usepackage{xcolor}
\usepackage{slashbox}
\begin{document}

\preprint{APS/123-QED}

\title{The robust-yet-fragile nature of interdependent networks}

\author{Fei Tan}
\author{Yongxiang Xia}
\email{xiayx@zju.edu.cn}

\affiliation{Department of Information Science and Electronic Engineering, Zhejiang University, Hangzhou 310027, China
}

\date{\today}% It is always \today, today,

\begin{abstract}
Interdependent networks have been shown to be extremely vulnerable based on the percolation model. Parshani et. al further
indicated that the more inter-similar networks are, the more robust they are to random failure. Our understanding of how
coupling patterns shape and impact the cascading failures of loads in interdependent networks is limited, but is essential
for the design and optimization of the real-world interdependent networked systems. This question, however, is largely unexplored.
In this paper, we address this question by investigating the robustness of interdependent ER random graphs and BA scale-free networks
under both random failure and intentional attack. It is found that interdependent ER random graphs are robust-yet-fragile under both
random failures and intentional attack. Interdependent BA scale-free networks, however, are only robust-yet-fragile under random failure
but fragile under intentional attack. These results advance our understanding of the robustness of interdependent networks
significantly.

\begin{description}
\item[DOI]  \item[PACS number(s)]
89.75.Hc, 89.75.Fb, 05.10.-a, 89.40.-a
\end{description}
\end{abstract}

\pacs{Valid PACS appear here}% PACS, the Physics and Astronomy
\maketitle

%\tableofcontents

\section{\label{sec:level1}Introduction}
We live in a modern society supported by many critical networked infrastructures such as power grids, communication and transportation
systems \cite{cui2010complex}. The robustness of systems has been undoubtedly the central issue \cite{albert2000error,watts2002simple,buldyrev2010catastrophic}.
Over the past decade, cascades on complex networks have thus been widely explored \cite{motter2002cascade,xia2010cascading}.
It has been revealed that scale-free networks are robust to random failure but fragile to
intentional attacks, whereas the random graph is robust to both \cite{motter2002cascade}. Xia et al. further indicated that the robust-yet-fragile (RYF) property
in cascades scenario is principally associated with the heterogeneity of network betweenness distribution rather than degree distribution \cite{xia2010cascading}.
Small-world networks, for example, are also robust-yet-fragile although its degree distribution is homogenous.

The aforementioned conclusions, however, are mainly derived based on the framework of isolated networks. Actually, various networked
infrastructures interact with each other significantly. Some coupled network models have been thus established to further capture
real-world networked systems \cite{buldyrev2010catastrophic,brummitt2012suppressing,gao2011networks}. Recently, interdependent networks have been formalized to explore the robustness of the systems in which
two different subsystems have to depend on each other \cite{buldyrev2010catastrophic}. In other words, the failure of nodes in one network will trigger the
counterpart ones in the other network to collapse accordingly. Buldyrev et al. has showed that such dependency between networks significantly increases the
vulnerability of networks
so that interdependent networks are even vulnerable to random failure. Based on the percolation theory,
deeper studies have focused on the effect of coupling pattern \cite{parshani2010inter}, coupling strength \cite{dong2012percolation,parshani2010interdependent} and network structures \cite{gao2011robustness} on the robustness of interdependent
networks. Such conceptual breakthrough shift from the isolation towards the interaction allows us to understand real-world networks more
readily. As another form of interaction, interconnected networks have also been developed to capture the coupling between two originally
isolated networks \cite{brummitt2012suppressing,tan2013cascading}. Different from dependency links between two parts for interdependent networks, the interconnected links across two
networks play as the same role as the connectivity within networks. For cascades of loads in interconnected networks, the optimal coupling
patten and/or coupling probability could be found.

Recently, traffic overload has been extended to the cascades on interdependent networks. Undoubtedly, such dependency will aggravate cascading
failures of loads in interdependent networks \cite{zhang2013robustness}. Nevertheless, is the extent to which network robustness against cascades of loads is
undermined by the dependency relation always large enough? In this paper, we will try to answer this question by considering network structures,
attack strategies and the coupling mechanisms.

\section{\label{sec:level2} Models}

\subsection{Network model}

For simplicity and clarity of the results, our model considers two networks A and B with the same size ($N=N_A=N_B$) and same
average degree($\langle k \rangle = \langle k_A \rangle = \langle k_B \rangle$). Here networks A and B are assumed to fully coupled,
and each node has only one interdependent link. Different from interconnected networks, the coupling links just refer to the dependency
relationships between two networks. As mentioned in previous studies, the way in which the coupling links are established has great impact
on the robustness of coupled networks. As we focus on how traffic overload leads to cascading failures, three coupling patterns are thus
described based on load distribution in individual networks \cite{tan2013cascading}.

\begin{itemize}
  \item  Assortative Coupling. Nodes are first sorted in networks A and B respectively, both in the descending order of load. If different
  nodes share the same load, we sort them at random. Connect the first node in network A with the first node in network B, and then connect
  the second node in network A with the second node in network B, and so on. Repeat this process until all interdependent links are built.

  \item Disassortative Coupling. Nodes are first sorted in network A (B) in the descending (ascending) order of load.  If different nodes
  share the same load, we sort them at random. Connect the first node in network A (with the heaviest load) with the first node in
  network B (with the lightest load), and then connect the second node in network A with the second node in network B, and so on.
  Repeat this process until all interdependent links are built.

  \item Random Coupling. Randomly choose a node in network A and a node in network B. If neither of them has an interdependent link,
  then connect them. Repeat this process until all interdependent links are built.
\end{itemize}

In order to maintain tractability and facilitate comparison with isolated complex networks, we presume that networks A and B share
the same type of networks. In this paper, we mainly focus on the robustness of coupled ER random graphs and coupled BA scale-free
networks.

\subsection{Traffic model}
In this paper, we adopt the data-packet transport scenario based on the shortest-path routing \cite{tan2013cascading,motter2002cascade}.
The node betweenness can thus approximate the traffic load. To be concrete, the betweenness of node $k$ is denoted as \cite{freeman1977set}
\begin{equation}
\label{betweenness}
B_k=\sum_{s\neq k \neq t}\frac{n_{st}^k}{g_{st}},
\end{equation}
where $g_{st}$ is the total number of possible shortest topological paths from node $s$ to node $t$ and $n_{st}^k$ denotes the number
of such paths running through node $k$. $n_{st}^k / g_{st}$ is stipulated to be zero, if $n_{st}^k = g_{st} = 0$.

The capacity of a node is the maximum load that the node can handle. The capacity of node $k$ is thus set to be proportional to its
initial load $L_k$ \cite{motter2002cascade,xia2010cascading,wang2009cascade}
\begin{equation}
C_{i}=(1+\alpha)L_{k},
\end{equation}
where the constant $\alpha$ is the tolerance parameter. The network is in the free-flow state (without overload) if $\alpha \geq 0$.

When one node is removed from the network (due to a failure or attack), it will trigger two kinds of effects.
On the one hand, the removal affects the shortest paths between some other nodes within the attacked network (e.g., network A). Then the
packets from these nodes have to adjust their paths. In this way, loads of many nodes are changed. Those nodes with loads greater than
their capacities will fail, then it triggers further load adjustment over and over until loads of all remaining nodes are less than their
capacities. It is obvious that the tolerance parameter $\alpha$ affects the effect of cascading failures.
On the other hand, the removed node in network A will cause the dependent counterpart node in network B to fail. The failure of this node
will cause the same cascades within network B. Actually, these two effects interact with each other in cascading failures. Note that
nodes outside the giant component are assumed to be failed. Therefore, the dependency relationship makes the cascading failure process
more complicated compared to isolated networks.

In this paper, two attack strategies will be explored, i.e., random failure and intentional attack. We use the relative size of the
giant component $G$ to characterize the robustness of networks. In our model, the remaining giant component after cascading failures in
two subnetworks share the same size. Thus, $G$ is the ratio between the remaining giant component size and original network size.

\section{\label{sec:level3}Results}

\begin{figure}[Ht!]
\centering
\includegraphics[height=\columnwidth,width=\columnwidth]{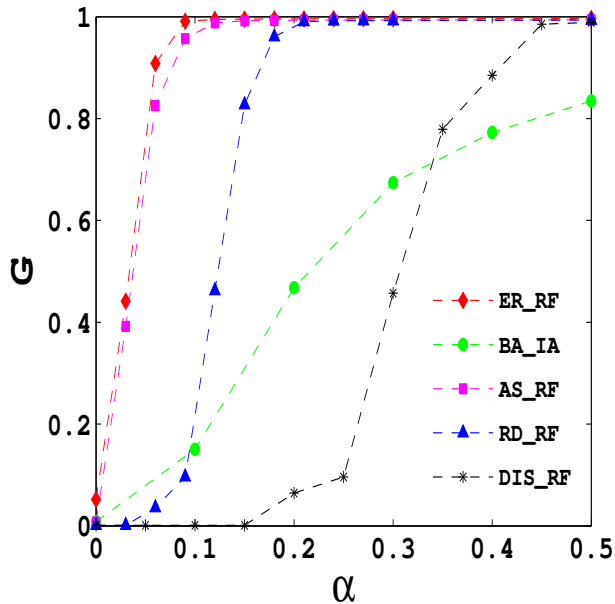}
\caption{(Color online) Cascading failures of coupled ER random graphs under random failure. The network size $N_A = N_B = 1000$ and average
degree $\langle k_A \rangle = \langle k_B \rangle = 6$. The relative size of the giant component $G$ is shown as a function of the
tolerance parameter $\alpha$ with different types of coupling patterns, namely, assortative coupling (\textcolor{magenta}{$\blacksquare$}),
random coupling (\textcolor{blue}{$\blacktriangle$}) and disassortative coupling ($\ast$). Cascading failures of the single ER random graph under
random failure (\textcolor{red}{$\blacklozenge$}) and the single BA scale-free network under intentional attack (\textcolor{green}{$\bullet$}).}
\label{ER_RF}
\end{figure}

\begin{figure}[Ht!]
\centering
\includegraphics[height=\columnwidth,width=\columnwidth]{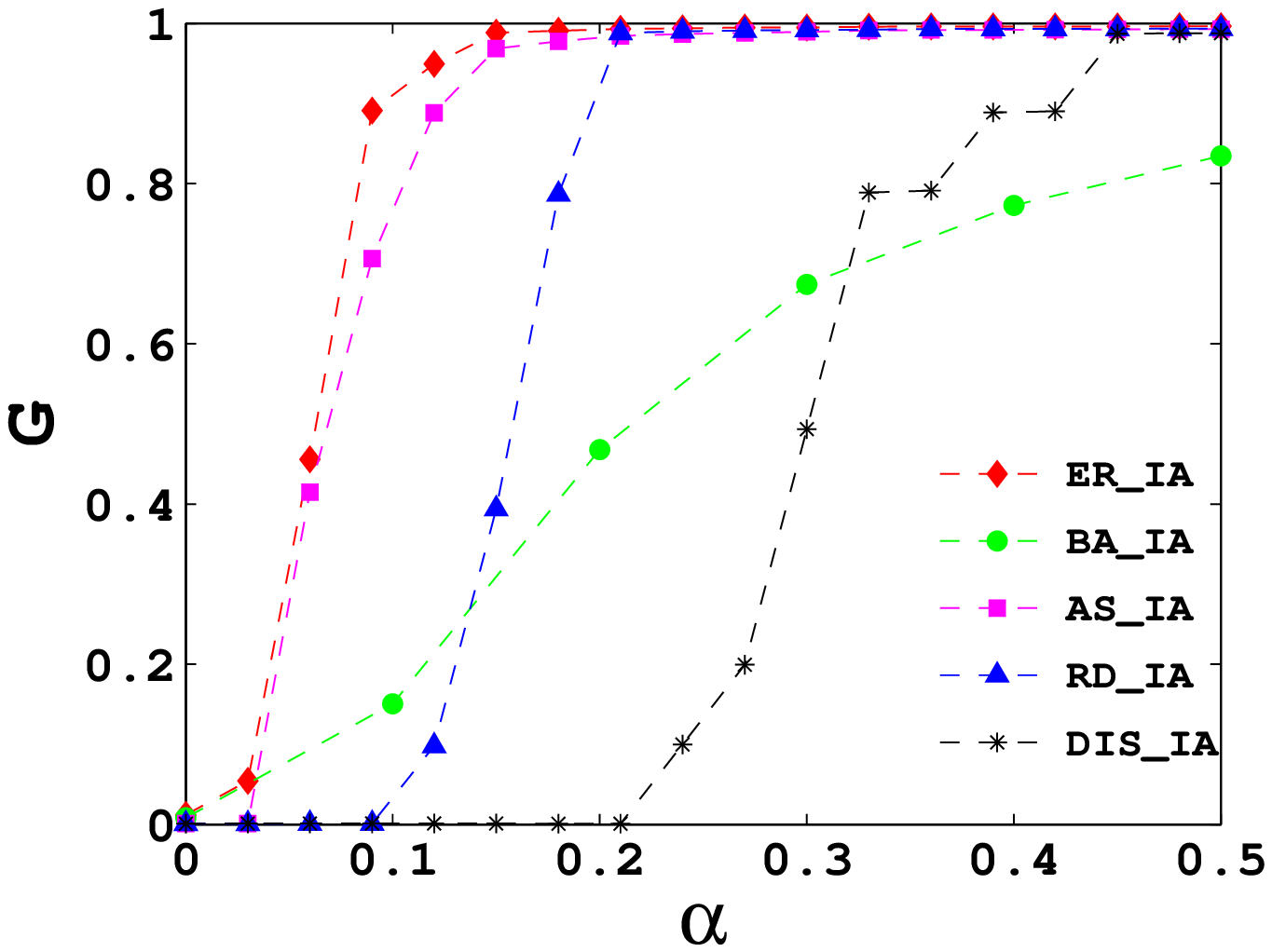}
\caption{(Color online) Cascading failures of coupled ER random graphs under intentional attack. The network size $N_A = N_B = 1000$ and
average degree $\langle k_A \rangle = \langle k_B \rangle = 6$. The relative size of the giant component $G$ is shown as a function
of the tolerance parameter $\alpha$ with different types of coupling patterns, namely, assortative coupling
(\textcolor{magenta}{$\blacksquare$}), random coupling (\textcolor{blue}{$\blacktriangle$}) and disassortative coupling ($\ast$).
Cascading failures of the single ER random graph (\textcolor{red}{$\blacklozenge$}) and BA scale-free network (\textcolor{green}{$\bullet$})
under intentional attack.}
\label{ER_IA}
\end{figure}

\begin{figure}[Ht!]
\centering
\includegraphics[height=\columnwidth,width=\columnwidth]{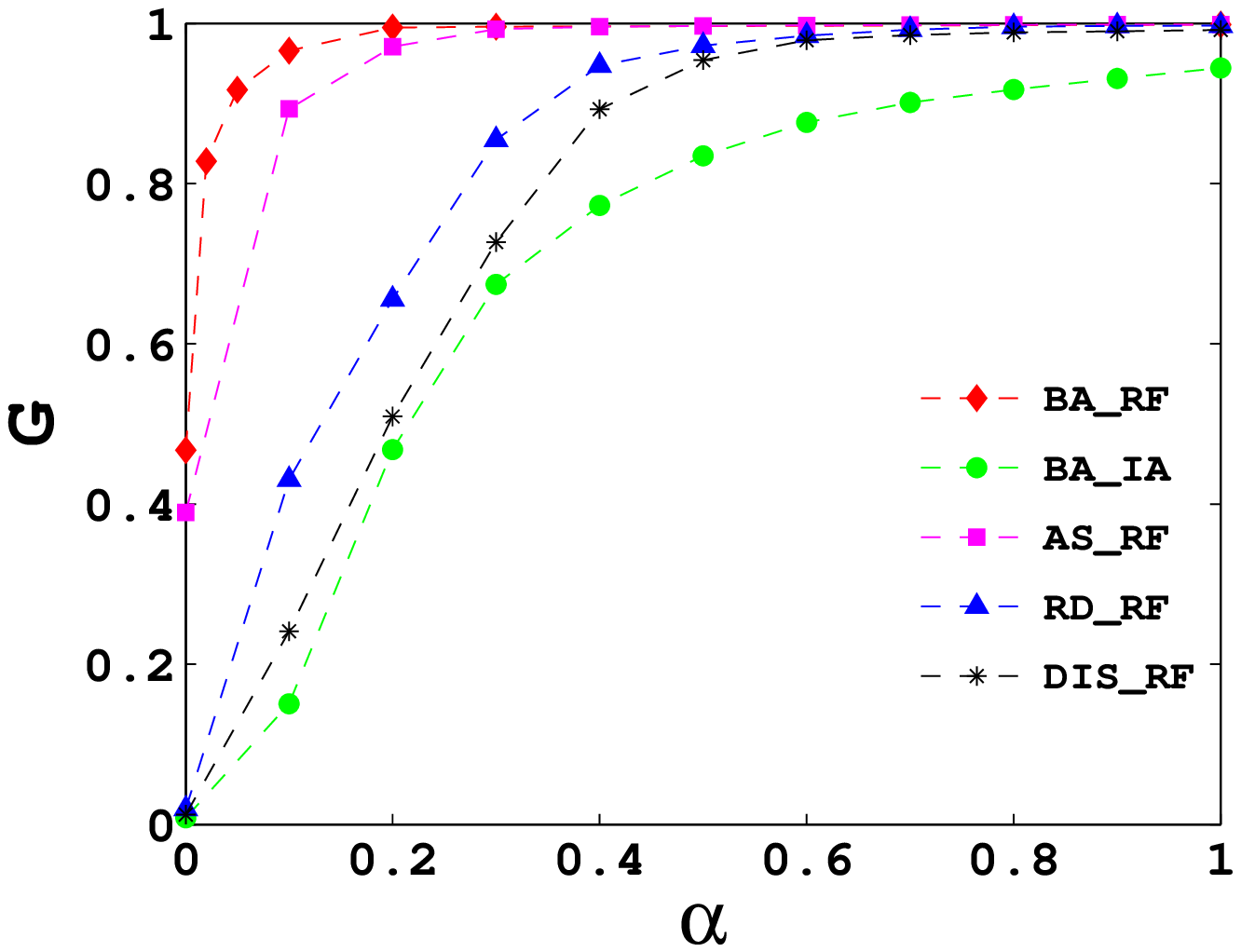}
\caption{(Color online) Cascading failures of coupled BA scale-free networks under random failure. The network size $N_A = N_B = 1000$ and
average degree $\langle k_A \rangle = \langle k_B \rangle = 6$. The relative size of the giant component $G$ is shown as a function of the
tolerance parameter $\alpha$ with different types of coupling patterns, namely, assortative coupling (\textcolor{magenta}{$\blacksquare$}),
random coupling (\textcolor{blue}{$\blacktriangle$}) and disassortative coupling ($\ast$). Cascading failures of the single BA scale-free
network under random failure (\textcolor{red}{$\blacklozenge$}) and intentional attack (\textcolor{green}{$\bullet$}). }
\label{BA_RF}
\end{figure}

\begin{figure}[Ht!]
\centering
\includegraphics[height=\columnwidth,width=\columnwidth]{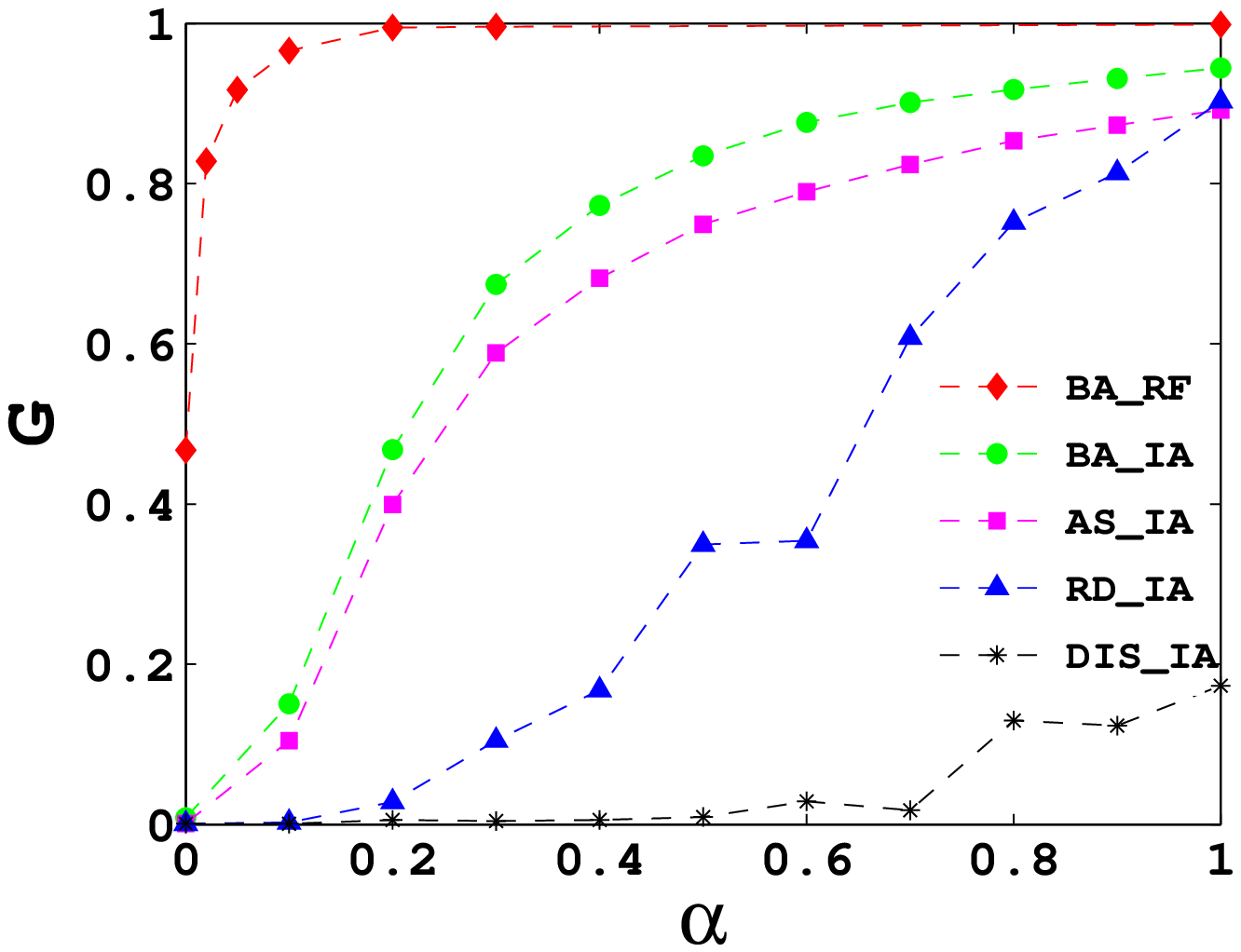}
\caption{(Color online) Cascading failures of coupled BA scale-free networks under intentional attack. The network size $N_A = N_B = 1000$ and
average degree $\langle k_A \rangle = \langle k_B \rangle = 6$. The relative size of the giant component $G$ is shown as a function of the
tolerance parameter $\alpha$ with different types of coupling patterns, namely, assortative coupling (\textcolor{magenta}{$\blacksquare$}),
random coupling (\textcolor{blue}{$\blacktriangle$}) and disassortative coupling ($\ast$). Cascading failures of the single BA scale-free
network under random failure (\textcolor{red}{$\blacklozenge$}) and intentional attack (\textcolor{green}{$\bullet$}). }
\label{BA_IA}
\end{figure}

In this section, we will carry out numerical simulations on coupled ER random graphs and coupled BA scale-free networks.
We mainly explore four kinds of situations, i.e., (1) coupled ER random graphs under random failure, (2) coupled ER random graphs under 
intentional attack, (3) coupled BA scale-free networks under random failure, (4) coupled BA scale-free networks under intentional attack.

We first investigate the robustness of interdependent ER random graphs under random failure. As ER random graph is robust to random
failure, and BA scale-free network is fragile to intentional attack, we view them as the baselines. First,
we find that the dependency relationship between two networks reduces the network robustness. As shown in fig. 1, under random failure,
the robustness of interdependent networks under three coupling mechanisms is worse than that of the single networks. Second, the
coupling patterns influence the robustness of interdependent networks remarkably.
It is shown that the assortative $>$ the random coupling $>$ the disassortative coupling, in terms of network robustness.
Note that the robustness of interdependent random graphs with assortative coupling approaches that of the single ER random graph under
random failure. Interdependent ER random graphs with the assortative coupling can thus be
regarded as robust. Nevertheless, it is obvious that the disassortative coupling is fragile to random failure. They are even more 
vulnerable than the single BA scale-free networks under intentional attack. Consider, for example,
the tolerance $\alpha = 0.25$ in which the relative size of the giant component $G < 0.2$. When $\alpha$ reaches $0.4$, $G$ is still less
than $0.9$. Altogether, when random failure occurs, interdependent ER random graphs are robust with the assortative coupling but fragile with
disassortative coupling.

In situation (2), the robustness of interdependent ER random graphs under intentional attack is explored. As ER random graph is robust
to intentional attack, and BA scale-free network is fragile to intentional attack, we view them as the baselines. First, we find that 
the dependency relationship between two networks reduces the network robustness. As shown in fig. 2, under intentional attack, the 
robustness of interdependent ER random graphs under three
coupling mechanisms is worse than that of the single ER random graph. Second, the coupling patterns influence the robustness of interdependent
networks remarkably. It is shown that the assortative $>$ the random coupling $>$ the disassortative coupling, in terms of network
robustness. Note that the robustness of the interdependent network with assortative coupling approaches that of the single network under
intentional attack. Interdependent ER random graphs with the assortative coupling can
thus be regarded as robust. Nevertheless, it is obvious that the disassortative coupling is fragile to intentional attack. They are even more
vulnerable than the single BA scale-free networks under intentional attack. Consider, for
example, the tolerance $\alpha = 0.2$ in which coupled networks are still fragmented. When $\alpha$ reaches $0.4$, $G$ is still less than
$0.9$. Thus, similarly to random failure, interdependent ER random graphs remains robust-yet-fragile under intentional attack.

Take the above two situations together, under both random failure and intentional attack, interdependent ER random graphs are robust for the
assortative coupling but fragile for the disassortative coupling. Thus, interdependent ER random graphs are robust-yet-fragile.

We further explore the robustness of interdependent BA scale-free networks under random
failures. As BA scale-free network is robust to random failure but fragile to intentional attack, we view them as the baselines.
First, we find that the dependency relationship between two networks reduces the network robustness. As shown in fig. 3, under random failure, 
the robustness of interdependent BA scale-free
networks under three coupling mechanisms is worse than that of the single BA scale-free network. Second, the coupling patterns influence the
robustness of interdependent networks remarkably. It is shown that the assortative $>$ the random coupling $>$ the disassortative coupling,
in terms of network robustness. Note that the robustness of interdependent BA scale-free networks with assortative coupling approaches that of the
single BA scale-free network under random failure. The $G-\alpha$ curve of the disassortative coupling, however, is close to that of the single
BA scale-free network under intentional network. Interdependent BA scale-free networks can thus be regarded as robust-yet-fragile.

\begin{table}[!t]
\renewcommand{\arraystretch}{1.2}
\caption{The robustness of different network models under random failure and intentional attack.}

\label{Attack}
\centering
\begin{tabular}{|p{2.8cm}<{\centering}|p{2.6cm}<{\centering}|p{2.6cm}<{\centering}|}
\hline
\backslashbox{Network}{Attack} & Random Failure   & Intentional Attack   \\
\hline
ER  & Robust & Robust \\
\hline
BA & Robust & Fragile \\
\hline
\end{tabular}
\end{table}

\begin{table}[!t]
\renewcommand{\arraystretch}{1.2}
\caption{The robustness of Interdependent ER networks with different coupling patterns under random failure and intentional attack.}
\label{ER}
\centering
\begin{tabular}{|p{2.8cm}<{\centering}|p{2.6cm}<{\centering}|p{2.6cm}<{\centering}|}
\hline
\backslashbox{Coupling}{Attack} & Random Failure & Intentional Attack\\
\hline
Assortative & Robust & Robust \\
\hline
Disassortative & Fragile & Fragile \\
\hline
\end{tabular}
\end{table}

\begin{table}[!t]
\renewcommand{\arraystretch}{1.2}
\caption{The robustness of Interdependent scale-free networks with different coupling patterns under random failure and intentional attack.}
\label{BA}
\centering
\begin{tabular}{|p{2.8cm}<{\centering}|p{2.6cm}<{\centering}|p{2.6cm}<{\centering}|}
\hline
\backslashbox{Coupling}{Attack} & Random Failure & Intentional Attack\\
\hline
Assortative & Robust & Fragile \\
\hline
Disassortative & Fragile & Fragile \\
\hline
\end{tabular}
\end{table}

Similarly to the baselines of situation (3), we lastly explore the robustness of interdependent scale-free networks under
intentional attack. First, we find that the dependency relationship between two networks reduces the network robustness. As shown in
fig. 4, under intentional attack, the robustness of interdependent BA scale-free networks under three coupling mechanisms is worse than
that of the single BA scale-free network. Second, the coupling patterns influence the robustness of interdependent networks remarkably.
It is shown that the assortative $>$ the random coupling $>$ the disassortative coupling, in terms of network robustness. Take the
tolerance parameter $\alpha = 0.7$ for example, the relative size of the giant component of the assortative coupling reaches $0.8$,
whereas it is about zero for the assortative coupling. Interdependent scale-free networks can thus be regarded as fragile.

Collecting these four situations (as shown in the Tables \ref{Attack}, \ref{ER} and \ref{BA}), we find that interdependent ER random graphs
are robust-yet-fragile under both random failure and intentional attack. Interdependent BA scale-free networks are robust-yet-fragile under
random failure but fragile under intentional attack. Note that these results are different from the extreme vulnerability of interdependent
networks based on the percolation theory.

\section{\label{sec:level4}Conclusions}
In this paper, we have explored cascades of loads in interdependent networks. The interplay of traffic overload and dependency makes
the robustness of interdependent networks more complicated than the single network. For interdependent ER random graphs under
both random failure and intentional attack, they are robust for the assortative coupling but fragile for the disassortative coupling. For
interdependent BA scale-free networks, under random failure, they are robust for the assortative coupling but fragile for the disassortative
coupling, whereas under intentional attack, they are fragile for both. Theoretically, the robust-yet-fragile property can refresh our
understanding of the robustness of interdependent networks. Furthermore, our results can provide useful insights for the design and
optimization of real-world interdependent networked systems.

\begin{acknowledgments}
This work was supported by the National Natural Science Foundation of China under Grant No. 61174153.
\end{acknowledgments}

%\nocite{*}

\bibliography{Cascading_failure}% Produces the bibliography via BibTeX.

\end{document}